# Relativistic five-quark equations and exotic baryons with isospin I = 5/2

Gerasyuta S.M. [1,2], Kochkin V.I. [1]

1. Department of Theoretical Physics, St. Petersburg State University, 198904, St. Petersburg, Russia.
2. Department of Physics, LTA, 194021 St. Petersburg, Russia.

Abstract.

The relativistic five-quark equations are found in the framework of the dispersion relation technique. The five-quark amplitudes for the low-lying exotic baryons including the $u, d$ quarks are calculated. The poles of the five-quark amplitudes determine the masses and the widths of $uuu\bar{d}$ and $ddd\bar{u}$ states (isospin I = 5/2). The mass spectrum of the $E^{+++}$ exotic baryons with $J^P = \frac{1}{2}^+, \frac{3}{2}^+, \frac{5}{2}^+$ is calculated.

PACS numbers: 11.55 Fv, 12.39 Ki, 12.39 Mk, 12.40 Yx.

I. Introduction

The existence of particles made of more than three quarks is an important issue in QCD-inspired models. The several experimental groups [1-10] have reported observation of a new exotic (S = 1) baryon resonance $\theta^+$(1540) with a narrow decay width. Recently, the CLAS experiment has reported null results of finding the $\theta^+$ in the $\gamma p \to \overline{K}_0 \theta^+$ reaction [11]. Though this experiment has high statistics, the result does not lead to the absence of $\theta^+$ immediately, because the previous positive evidences were seen mostly in the reaction from the neutron.

The chiral soliton or Skyrme model has played an important role in stimulating the search for exotics [12-16]. Diakonov et al [17] have studied the properties of $\theta^+$ as a member of the antidecuplet, with prediction of mass and width close to the results of experimentalists.

Since the $\theta^+$ discovery, there has been a flurry of papers on pentaquarks in the constituent quark model [18-21], correlated quark model [22, 23], chiral soliton model [24-27]. In the Refs. [28-32], a method has been developed which is convenient for analyzing relativistic three-hadron systems. The physics of three-hadron systems can be described by means of a pair interaction between the particles. There are three isobar channels, each of which consists of a two-particle isobar and the third particle. The presence of the isobar representation together with the condition of unitarity in the pair energies and of analyticity leads to a system of integral equations in a single variable. Their solution makes it possible to describe the interaction of the produced particles in a three-hadron system.

In our papers [33, 34] relativistic generalization of the three-body Faddeev equations was obtained in the form of dispersion relations in the pair energy of two interacting particles. The mass spectrum of S-wave baryons including u, d, s – quarks was calculated by a method based on isolating the leading singularities in the amplitude. We searched for the approximate solution of integral three-quark equations by taking into account two-particle and triangle singularities, all the weaker ones being neglected. If we considered such an approximation, which corresponds to taking into account two-body and triangle singularities, and defined all the smooth functions of the subenergy variables (as compared with the singular part of the amplitude) in the middle point of the physical region of Dalitz-plot, then the problem was reduced to the one of solving a system of simple algebraic equations.

In our previous papers [35, 36] the relativistic five-quark equations for the family of the $\theta$ pentaquarks are constructed. The five-quark amplitudes for the low-lying $\theta$ pentaquarks are calculated. The mass of $\theta^+$ pentaquark with positive parity is found to be smaller than the mass of $\theta^+$ pentaquark with negative parity. Our calculation take into account the contributions to the $\theta^+$ amplitude not only for 8-,10*-plets ,but also for the 35-, 27-, 10-plets of $SU(3)_f$.

The present paper is devoted to the construction of relativistic five-quark equations for two flavors (u, d) exotic baryons. The five-quark amplitudes for the lowest exotic baryons contain only light quarks $uuuu\bar{d}$ and $dddd\bar{u}$ (isospin I = 5/2). The exotic baryon was considered with the antiquark in the P-wave. The poles of these amplitudes determine the masses and the widths of exotic baryons. The constituent quark is the color triplet and the quark amplitudes obey the global color symmetry. The interesting result of this model is the calculation of the exotic baryon amplitudes which contain the contribution of four subamplitudes: molecular subamplitude $BM$, $D\bar{q}D$ subamplitude, $Mqqq$ subamplitude and $Dqq\bar{q}$ subamplitude. Here $B$ corresponds to the lowest baryon with diquark $J^P = 1^+$. $M$ are the low-lying mesons with the quantum numbers: $J^{PC} = 0^{++}, 1^{++}, 2^{++}$. The masses of the constituent u, d-quarks coincide with the quark mass of the ordinary baryons in our quark model [33] m = 410 MeV. The model has only three parameters. The cutoff parameters $\Lambda_{0^+}$ = 16.5 and $\Lambda_{1^+}$ = 20.12, the gluon coupling constant $g$ = 0.417 are similar to Ref. [37].

The paper is organized as follows. After this introduction, we discuss the five-quark amplitudes, which consist of the u, d quarks (Sec. 2). In Section 3 we report our numerical results (Table).

II. Exotic baryon amplitudes.

We derived the relativistic five-quark equations in the framework of the dispersion relation technique. We use only planar diagrams; the other diagrams due to the rules of $1/N_c$ expansion [38-40] are neglected. The correct equations for the amplitude are obtained by taking into account all possible subamplitudes. It corresponds to the division of the complete system into subsystems with a smaller number of particles. Then one should represent a five-particle amplitude as a sum of ten subamplitudes:

$$A = A_{12} + A_{13} + A_{14} + A_{15} + A_{23} + A_{24} + A_{25} + A_{34} + A_{35} + A_{45}. \tag{1}$$

This defines the division of the diagrams into group according to certain pair interaction of particles. The total amplitude can be represented graphically as a sum of diagrams.

We need to consider only one group of diagrams and the amplitude corresponding to them, for example $A_{12}$. We shall consider the derivation of the relativistic equations of the Faddeev-Yakubovsky approach [41, 42] for the exotic baryon. We shall construct the five-quark amplitude of four $u$ quarks and one $d$ antiquark, in which the pair interaction with the quantum number of $1^+$ is included. The set of diagrams associated with the amplitude $A_{12}$ can be further broken down into four groups corresponding to amplitudes $A_1(s, s_{1234}, s_{12}, s_{34})$, $A_2(s, s_{1234}, s_{25}, s_{34})$, $A_3(s, s_{1234}, s_{13}, s_{134})$, $A_4(s, s_{1234}, s_{23}, s_{234})$ (Fig.). The antiquark is shown by the arrow and the other lines correspond to the quarks. The coefficients are determined by the permutation of quarks [41, 42]. Here $s_{ik}$ is the two-particle subenergy squared, $s_{ijk}$ corresponds to the energy squared of particles $i$, $j$, $k$, $s_{ijkl}$ is the four-particle subenergy squared and $s$ is the system total energy squared.

In order to represent the subamplitudes $A_1(s, s_{1234}, s_{12}, s_{34})$, $A_2(s, s_{1234}, s_{25}, s_{34})$, $A_3(s, s_{1234}, s_{13}, s_{134})$, and $A_4(s, s_{1234}, s_{23}, s_{234})$ in the form of a dispersion relation, it is necessary to define the amplitudes of quark-quark and quark-antiquark interaction $b_n(s_{ik})$. The pair quark amplitudes $q\bar{q} \to q\bar{q}$ and $qq \to qq$ are calculated in the framework of the dispersion N/D method with the input four-fermion interaction [43, 44] with quantum numbers of the gluon [45]. The regularization of the dispersion integral for the D-function is carried out with the cutoff parameters $\Lambda_n$.

The four-quark interaction is considered as an input [45]:

$$g_v (\bar{q} \vec{\lambda} I_f \gamma_\mu q)^2 \tag{2}$$

Here $I_f$ is the unity matrix in the flavor space (u, d), $\vec{\lambda}$ are the color Gell-Mann matrices. Dimensional constant of the four-fermion interaction $g_v$ is the parameter of the model.

Dimensionless parameters $g$ and $\Lambda_n$ are supposed to be constants which are independent of the quark interaction type. The applicability of Eq. (2) is verified by the success of De Rujula-Georgi-Glashow quark model [46], where only the short-range part of Breit potential connected with the gluon exchange is responsible for the mass splitting in hadron multiplets.

We use the results of our relativistic quark model [45] and write down the pair quarks amplitude in the form:

$$b_n(s_{ik}) = \frac{G_n^2(s_{ik})}{1 - B_n(s_{ik})}, \quad (3)$$

$$B_n(s_{ik}) = \int_{4m^2}^{\Lambda} \frac{ds_{ik}'}{\pi} \frac{\rho_n(s_{ik}')G_n^2(s_{ik}')}{s_{ik}' - s_{ik}}. \quad (4)$$

$G_n(s_{ik})$ are the quark-quark and quark-antiquark vertex functions. The vertex functions satisfy the Fierz relations. All of these vertex functions are generated from $g_v$. $B_n(s_{ik})$, $\rho_n(s_{ik})$ are the Chew-Mandelstam functions with the cutoff $\Lambda_n$ and the phase spaces respectively. Here n=1 describes a $qq$-pair with $J^P = 1^+$ in the $\bar{3}_c$ color state and n=2 defines the $q\bar{q}$-pairs, corresponding to mesons with quantum numbers: $J^{PC} = 0^{++}, 1^{++}, 2^{++}$.

In the case in question the interacting quarks do not produce a bound state; therefore the integration in Eqs.(5) - (8) below is carried out from the threshold $4m^2$ to the cutoff $\Lambda_n$. The system of integral equations, corresponding to the meson state with $J^{PC} = 0^{++}$ and diquark with $J^P = 1^+$, can be described as:

$$A_1(s, s_{1234}, s_{12}, s_{34}) = \frac{\lambda_1 B_2(s_{12})B_1(s_{34})}{[1 - B_2(s_{12})][1 - B_1(s_{34})]} + 6\hat{J}_2(2,1)A_4(s, s_{1234}, s_{23}', s_{234}') +$$
$$+ 2\hat{J}_2(2,1)A_3(s, s_{1234}, s_{13}', s_{134}') + 2\hat{J}_1(2)A_3(s, s_{1234}, s_{15}', s_{125}') + 2\hat{J}_1(2)A_4(s, s_{1234}, s_{25}', s_{125}') +, \quad (5)$$
$$+ 4\hat{J}_1(1)A_4(s, s_{1234}, s_{35}', s_{345})$$

$$A_2(s, s_{1234}, s_{25}, s_{34}) = \frac{\lambda_2 B_1(s_{25})B_1(s_{34})}{[1 - B_1(s_{25})][1 - B_1(s_{34})]} + 12\hat{J}_2(1,1)A_4(s, s_{1234}, s_{23}', s_{234}') +$$
$$+ 8\hat{J}_1(1)A_3(s, s_{1234}, s_{12}', s_{125}) \quad (6)$$

$$A_3(s, s_{1234}, s_{13}, s_{134}) = \frac{\lambda_3 B_2(s_{13})}{1 - B_2(s_{13})} + 12\hat{J}_3(2)A_1(s, s_{1234}, s_{12}', s_{34}'), \quad (7)$$

$$A_4(s,s_{1234},s_{23},s_{234}) = \frac{\lambda_4 B_1(s_{23})}{1-B_1(s_{23})} + 4\hat{J}_3(1)A_2(s,s_{1234},s'_{25},s'_{34}) + 4\hat{J}_3(1)A_1(s,s_{1234},s'_{12},s'_{34}), \quad (8)$$

were $\lambda_i$ are the current constants. We introduced the integral operators:

$$\hat{J}_1(l) = \frac{G_l(s_{12})}{[1-B_l(s_{12})]} \int_{4m^2}^{\Lambda} \frac{ds'_{12}}{\pi} \frac{G_l(s'_{12})\rho_l(s'_{12})}{s'_{12}-s_{12}} \int_{-1}^{+1} \frac{dz_1}{2}, \quad (9)$$

$$\hat{J}_2(l,p) = \frac{G_l(s_{12})G_p(s_{34})}{[1-B_l(s_{12})][1-B_p(s_{34})]} \times$$
$$\times \int_{4m^2}^{\Lambda} \frac{ds'_{12}}{\pi} \frac{G_l(s'_{12})\rho_l(s'_{12})}{s'_{12}-s_{12}} \int_{4m^2}^{\Lambda} \frac{ds'_{34}}{\pi} \frac{G_p(s'_{34})\rho_p(s'_{34})}{s'_{34}-s_{34}} \int_{-1}^{+1} \frac{dz_3}{2} \int_{-1}^{+1} \frac{dz_4}{2}, \quad (10)$$

$$\hat{J}_3(l) = \frac{G_l(s_{12},\tilde{\Lambda})}{1-B_l(s_{12},\tilde{\Lambda})} \times$$
$$\times \frac{1}{4\pi} \int_{4m^2}^{\tilde{\Lambda}} \frac{ds'_{12}}{\pi} \frac{G_l(s'_{12},\tilde{\Lambda})\rho_l(s'_{12})}{s'_{12}-s_{12}} \int_{-1}^{+1} \frac{dz_1}{2} \int_{-1}^{+1} dz \int_{z_2^-}^{z_2^+} dz_2 \frac{1}{\sqrt{1-z^2-z_1^2-z_2^2+2zz_1z_2}}, \quad (11)$$

$l, p$ are equal 1 or 2. Here $m$ is a quark mass.

In Eqs.(9) and (11) $z_1$ is the cosine of the angle between the relative momentum of particles 1 and 2 in the intermediate state and the momentum of particle 3 in the final state, taken in the c.m. of particles 1 and 2. In Eq.(11) $z$ is the cosine of the angle between the momenta of particles 3 and 4 in the final state, taken in the c.m. of particles 1 and 2. $z_2$ is the cosine of the angle between the relative momentum of particles 1 and 2 in the intermediate state and the momentum of particle 4 in the final state, taken in the c.m. of particles 1 and 2. In Eq. (10): $z_3$ is the cosine of the angle between relative momentum of particles 1 and 2 in the intermediate state and the relative momentum of particles 3 and 4 in the intermediate state, taken in the c.m. of particles 1 and 2. $z_4$ is the cosine of the angle between the relative momentum of the particles 3 and 4 in the intermediate state and momentum of the particle 1 in the intermediate state, taken in the c.m. of particles 3 and 4.

We can pass from the integration over the cosines of the angles to the integration over the subenergies.

The solutions of the system of equations are considered as:

$$\alpha_i(s) = F_i(s,\lambda_i)/D(s), \quad (12)$$

where zeros of $D(s)$ determinants define the masses of bound states of exotic baryons. $F_i(s,\lambda_i)$ are the functions of $s$ and $\lambda_i$. The functions $F_i(s,\lambda_i)$ determine the contributions of subamplitudes to the exotic baryon amplitude.

### III. Calculation results.

In functions $F_i(s,\lambda_i)$ allow to obtain the overlap factors $f$ for the exotic baryons $E^{+++}$. We calculated the overlap factors and the phase spaces for the decay $E^{+++} \to \Delta^{++}\pi^+ \to p\pi^+\pi^+$.

Given typical width for baryons in the $E^{+++}$ mass range $\approx$ 200 MeV [47], we would estimate naively [48] the $E^{+++}$ widths. We calculated the masses and the widths of low-lying exotic baryons (Table). The results of calculations allow to consider two exotic baryons as narrow resonances. The width of $E^{+++}$, M = 1485 MeV, with the $J^P = \frac{1}{2}^+, \frac{3}{2}^+$ and isospin I = 5/2 is equal to 15 MeV. The width of $E^{+++}$, M = 1550 MeV, with the $J^P = \frac{1}{2}^+, \frac{3}{2}^+, \frac{5}{2}^+$, I = 5/2 is about 25 MeV. We predict the degenerace of $E^{+++}$ exotic baryons. Our calculations take into account the contributions to the $E^{+++}$ amplitude not only for 8,10*-plets but also 35-, 27-, 10-plets of $SU(3)_f$.

The exotic states $E^{+++}$ was considered in the papers [49, 50] for the various quark models and the experimental results. It is important to notice that two experimental groups [51, 52] have reported the observation of exotic baryons $E^{+++}$ (I = 5/2) with masses and widths: M=1.480$\pm$0.010 GeV, $\Gamma$=0.05$\pm$0.03 GeV; M=1.567$\pm$0.019 GeV, $\Gamma$=0.049$\pm$0.022 GeV; M=1.735$\pm$0.029 GeV, $\Gamma$=0.099$\pm$0.032 GeV. These groups have considered the reactions $\pi^+ p \to p\pi^+\pi^+\pi^-$ and $\pi^+ p \to p\pi^+\pi^+\pi^-\pi^0$ and have discussed the $E^{+++} \to \Delta^{++}(1232)\pi^+$ decay ($\pi^+$ in the P-wave). These results are in good agreement with the calculations (Table).

Acknowledgments.

The authors would like to thank T. Barnes, S. Chekanov, D.I. Diakonov, A. Hosaka, Fl. Stancu for useful discussions. This research was supported by the Russian Ministry of Education (grant 2.1.1.68.26).

Figure captions.

Fig. Graphic representation of the equations for the five-quark subamplitudes $A_1(s, s_{1234}, s_{12}, s_{34})$ $(BM)$, $A_2(s, s_{1234}, s_{25}, s_{34})$ $(D\bar{q}D)$, $A_3(s, s_{1234}, s_{13}, s_{134})$ $(Mqq)$, and $A_4(s, s_{1234}, s_{23}, s_{234})$ $(Dqq\bar{q})$ using the low-lying mesons with $J^{PC} = 0^{++}, 1^{++}, 2^{++}$ and diquark with $J^P = 1^+$.

Table. Low-lying exotic baryons $E^{+++}$ masses and widths.

| $E^{+++}$ | $J^P$ | Mass, MeV | Width (MeV) |
|---|---|---|---|
| $uuuu\bar{d}$ I=5/2 | $\frac{1}{2}^+, \frac{3}{2}^+$ | 1485 (1480 ± 10) | 15 (50 ± 30) |
| | $\frac{1}{2}^+, \frac{3}{2}^+, \frac{5}{2}^+$ | 1550 (1567 ± 19) | 25 (49 ± 22) |
| | $\frac{1}{2}^+, \frac{3}{2}^+, \frac{5}{2}^+$ | 1736 (1735 ± 29) | 160 (99 ± 32) |

Parameters of model: quark mass $m = 410$ MeV, cutoff parameter $\Lambda_{0^+} = 16.5$ and $\Lambda_{1^+} = 20.12$; gluon constant $g = 0.417$ [37]. Experimental mass and width values of exotic baryons $E^{+++}$ are given in parentheses [50-52].

.

References

1.LEPS Collaboration, T.Nakano et al.Phys.Rev.Lett.91, 012002(2003).

2.V.V.Barmin et al. Phys.Atom Nucl.66,1715(2003).

3.CLAS Collaboration, S.Stepanyan et al.Phys.Rev.Lett.91,252001(2003).

4.SAPHIR Collaboration, J.Barth et al.Phys.Lett.B572,127(2003).

5.A.E.Asratyan, A.G.Dolgolenko and M.A.Kubantsev,hep-ex/0309042.

6.HERMES Collaboration, A.Airapetian et al.Phys.Lett.B585,213(2004).

7.SVD Collaboration, A.Aleev et al.hep-ex/0401024.

8.COSY-TOF Collaboration, M.Abdel-Bary et al.hep-ex/0403011.

9.P.Zh.Aslanyan,V.N.Emelyanenko, G.G.Rikhvitzkaya.hep-ex/0403044.

10.ZEUS Collaboration, S.Chekanov et al.hep-ex/0403051.

11.CLAS Collaboration, R.De Vita et al. Talk given at APS April meeting(2005).

12.E.Guadagnini,Nucl.Phys.B236,35(1984).

13.A.V.Monohar,Nucl.Phys.B248,19(1984).

14.M.Chemtob,Nucl.Phys.B256,600(1985).

15.M.P.Mattis and M.Karliner,Phys.Rev.D31,2833(1985).

16.S.Jain and S.R.Wadia,Nucl.Phys.B258,713(1985).

17.D.Diakonov,V.Petrov and M.Polyakov,Z.Phys.A359,305(1997).

18.B.Jennings and K.Maltman,hep-ph/0308286.

19.C.E.Carlson,C.D.Carone,H.J.Kwee and V.Nazaryan.Phys.Lett.B573,101(2003).

20.F.Stancu and D.O.Riska,Phys.Lett.B575,242(2003).

21.S.M.Gerasyuta and V.I.Kochkin,Int.J.Mod.Phys.E12,793(2003).

22.R.J.Jaffe and F.Wilczek,Phys.Rev.Lett.91,23002(2003).

23.M.Karliner and H.J.Lipkin,Phys.Lett.B575,249(2003).

24.M.Praszalowitz,Phys.Lett.B583,96(2004).

25.J.Ellis,M.Karliner and M.Praszalowitz, J.High Energy Phys.0405,002(2004).

26.H.Weigel,Eur.Phys.J.A24,391(1998).

27.H.Walliser and V.B.Kopeliovich.J.Exp.Theor.Phys.97,433(2003),

28.I.J.R.Aitchison,J.Phys.G3,121(1977).

29.J.J.Brehm,Ann.Phys.(N.Y) 108,454(1977).

30.I.J.R.Aitchison and J.J.Brehm,Phys.Rev.D17,3072(1978).

31. I.J.R.Aitchison and J.J.Brehm,Phys.Rev.D20,1119(1979).

32. J.J.Brehm,Phys.Rev.D21,718(1980).

33. S.M.Gerasyuta,Sov.J.Nucl.Phys.55,3030(1992).

34. S.M.Gerasyuta,Z.Phys.C60,683(1993).

35. S.M.Gerasyuta andV.I.Kochkin,Phys.Rev.D71,076009(2005).

36. S.M.Gerasyuta and V.I.Kochkin,Phys.Rev.D72,016002(2005).

37. S.M.Gerasyuta and V.I.Kochkin,Int.J.Mod.Phys.E15,71(2006).

38. G.'t Hooft,Nucl.Phys.B72,461(1974).

39. G.Veneziano,Nucl.Phys.B117,519(1976).

40. E.Witten,Nucl.Phys.B160,57(1979).

41. O.A.Yakubovsky,Sov.J.Nucl.Phys.5,1312(1967).

42. S.P.Merkuriev and L.D.Faddeev, Quantum Scattering Theory for System of Few Particles (Nauka,Moscow,1985),p.398.

43. T.Appelqvist and J.D.Bjorken,Phys.Rev.D4,3726(1971).

44. C.C.Chiang,C.B.Chiu,E.C.G.Sudarshan and X.Tata,Phys.Rev.D25,1136(1982).

45. V.V.Anisovich,S.M.Gerasyuta and A.V.Saranstev,Int.J.Mod.Phys.A6,625(1991).

46 A.De Rujula, H.Georgi and S.L.Glashow,Phys.Rev.D12,147(1975).

47. C.E.Carlson,C.D.Carone,H.J.Kwee and V.Nazaryan,hep-ph/0312325.

48. J.J.Dudek and F.E.Close,hep-ph/0311258.

49. CLAS Collaboration, M.Ripani et al,hep-ex/0210054,hep-ex/0304034.

50. A.F.Nilov,Yad.Fiz.69,918(2006).

51. A.G.Drutskoi et al.Preprint ITEP-88-001.

52. A.V.Aref'ev et al.,Yad.Fiz.51,414(1990).